\documentclass[letterpaper,twocolumn,prl,aps,superscriptaddress,amsmath,amssymb,floatfix]{revtex4-1}
\usepackage{mathptmx}
\usepackage[latin9]{inputenc}
\usepackage{color}
\usepackage{verbatim}
\usepackage{float}
\usepackage{amsmath}
\usepackage{amssymb}
\usepackage{graphicx}
\usepackage{esint}

\renewcommand{\figurename}{\textbf{Fig.}}
\usepackage[unicode=true,
 bookmarks=true,bookmarksnumbered=false,bookmarksopen=false,
 breaklinks=false,pdfborder={0 0 1},backref=false,colorlinks=true]
 {hyperref}
\hypersetup{
 linkcolor=magenta,urlcolor=blue,citecolor=blue,pdfstartview={FitH},hyperfootnotes=false}

\makeatletter

\usepackage{textcomp}
\usepackage{epstopdf}

\pdfpageheight\paperheight
\pdfpagewidth\paperwidth

\@ifundefined{textcolor}{}{%
 \definecolor{BLACK}{gray}{0}
 \definecolor{WHITE}{gray}{1}
 \definecolor{RED}{rgb}{1,0,0}
 \definecolor{GREEN}{rgb}{0,1,0}
 \definecolor{BLUE}{rgb}{0,0,1}
 \definecolor{CYAN}{cmyk}{1,0,0,0}
 \definecolor{MAGENTA}{cmyk}{0,1,0,0}
 \definecolor{YELLOW}{cmyk}{0,0,1,0}
}

\usepackage{xcolor}
\usepackage{soul}
\newcommand{\ket}[1]{\ensuremath{\left|#1\right\rangle}}

\definecolor{blue}{rgb}{0,0,1}
\definecolor{red}{rgb}{1,0,0}
\definecolor{green}{rgb}{0,1,0}

\usepackage{soul}

\makeatother

\begin{document}

\title{Self-induced optical non-reciprocity}

\author{Zhu-Bo~Wang}
\thanks{These two authors contributed equally to this work.}
\affiliation{CAS Key Laboratory of Quantum Information \& CAS Center For Excellence in Quantum Information and Quantum Physics, University of Science and Technology of China, Hefei 230026, P. R. China.}

\author{Yan-Lei~Zhang}
\thanks{These two authors contributed equally to this work.}
\affiliation{CAS Key Laboratory of Quantum Information \& CAS Center For Excellence in Quantum Information and Quantum Physics, University of Science and Technology of China, Hefei 230026, P. R. China.}

\author{Xin-Xin~Hu}
\affiliation{CAS Key Laboratory of Quantum Information \& CAS Center For Excellence in Quantum Information and Quantum Physics, University of Science and Technology of China, Hefei 230026, P. R. China.}

\author{Guang-Jie~Chen}
\affiliation{CAS Key Laboratory of Quantum Information \& CAS Center For Excellence in Quantum Information and Quantum Physics, University of Science and Technology of China, Hefei 230026, P. R. China.}

\author{Ming~Li}
\affiliation{CAS Key Laboratory of Quantum Information \& CAS Center For Excellence in Quantum Information and Quantum Physics, University of Science and Technology of China, Hefei 230026, P. R. China.}

\author{Peng-Fei~Yang}
\affiliation{State Key Laboratory of Quantum Optics and Quantum Optics Devices, and Institute of Opto-Electronics, Shanxi University, Taiyuan 030006, China}
\affiliation{Collaborative Innovation Center of Extreme Optics, Shanxi University, Taiyuan 030006, China.}

\author{Xu-Bo~Zou}
\affiliation{CAS Key Laboratory of Quantum Information \& CAS Center For Excellence in Quantum Information and Quantum Physics, University of Science and Technology of China, Hefei 230026, P. R. China.}

\author{Peng-Fei~Zhang}
\email{zhangpengfei@sxu.edu.cn}
\affiliation{State Key Laboratory of Quantum Optics and Quantum Optics Devices, and Institute of Opto-Electronics, Shanxi University, Taiyuan 030006, China}
\affiliation{Collaborative Innovation Center of Extreme Optics, Shanxi University, Taiyuan 030006, China.}

\author{Chun-Hua~Dong}
\email{chunhua@ustc.edu.cn}
\affiliation{CAS Key Laboratory of Quantum Information \& CAS Center For Excellence in Quantum Information and Quantum Physics, University of Science and Technology of China, Hefei 230026, P. R. China.}

\author{Gang~Li}
\email{gangli@sxu.edu.cn}
\affiliation{State Key Laboratory of Quantum Optics and Quantum Optics Devices, and Institute of Opto-Electronics, Shanxi University, Taiyuan 030006, China}
\affiliation{Collaborative Innovation Center of Extreme Optics, Shanxi University, Taiyuan 030006, China.}

\author{Tian-Cai~Zhang}
\affiliation{State Key Laboratory of Quantum Optics and Quantum Optics Devices, and Institute of Opto-Electronics, Shanxi University, Taiyuan 030006, China}
\affiliation{Collaborative Innovation Center of Extreme Optics, Shanxi University, Taiyuan 030006, China.}

\author{Guang-Can~Guo}
\affiliation{CAS Key Laboratory of Quantum Information \& CAS Center For Excellence in Quantum Information and Quantum Physics, University of Science and Technology of China, Hefei 230026, P. R. China.}

\author{Chang-Ling~Zou}
\email{clzou321@ustc.edu.cn}
\affiliation{CAS Key Laboratory of Quantum Information \& CAS Center For Excellence in Quantum Information and Quantum Physics, University of Science and Technology of China, Hefei 230026, P. R. China.}
\affiliation{State Key Laboratory of Quantum Optics and Quantum Optics Devices, and Institute of Opto-Electronics, Shanxi University, Taiyuan 030006, China}

\date{\today}
\begin{abstract}
Non-reciprocal optical components are indispensable in optical applications, and their realization without any magnetic field arose increasing research interests in photonics~\cite{Asadchy2020,Tang2022}. Exciting experimental progress has been achieved by either introducing spatial-temporal modulation of the optical medium~\cite{Yu2009,Sounas2017} or combining Kerr-type optical nonlinearity with spatial asymmetry in photonic structures~\cite{Fan2012,Yang2019PRL,Yang2020}. However, extra driving fields are required for the first approach, while the isolation of noise and the transmission of the signal cannot be simultaneously achieved for the other approach. Here, we experimentally demonstrate a new concept of nonlinear non-reciprocal susceptibility for optical media and realize the completely passive isolation of optical signals without any external bias field. The self-induced isolation by the input signal is demonstrated with an extremely high isolation ratio of $63.4\,\mathrm{dB}$, a bandwidth of $2.1\,\mathrm{GHz}$ for $60\,\mathrm{dB}$ isolation, and a low insertion loss of around $1\,\mathrm{dB}$. Furthermore, novel functional optical devices are realized, including polarization purification and non-reciprocal leverage. The demonstrated nonlinear non-reciprocity provides a versatile tool to control light and deepen our understanding of light-matter interactions, and enables applications ranging from topological photonics~\cite{wang2009,Bahari2017} to unidirectional quantum information transfer in a network~\cite{Lodahl2017,Jiao2020}.
\end{abstract}
\maketitle

\noindent Optical non-reciprocity is of great importance due to the fundamental physics of light-matter interactions with broken time-reversal symmetry and their applications in non-reciprocal photonic devices~\cite{Asadchy2020,Sounas2017,Tang2022}. For instance, the non-reciprocity is highly desired in studies of topological photonic effects~\cite{wang2009,Bahari2017} and could also be applied to  unidirectional quantum information transfer in a network~\cite{Lodahl2017,Jiao2020}. Two well-known routes are widely utilized to break the reciprocity of the optical system, i.e., the magneto-optical medium under an external magnetic bias and nonlinear optics effects~\cite{Asadchy2020,Sounas2017}. However, regarding the practical difficulties accompanying magnetic fields and magneto-optic materials~\cite{Bi2011}, the nonmagnetic realization of non-reciprocity has aroused considerable research interest and has recently led to controversy~\cite{Jalas2013}.

%\noindent \textbf{One Sentence Summary}: The nonlinear non-reciprocal susceptibility of the optical medium is revealed and applied to high-performance optical isolation induced by the input signal itself.

It has been experimentally demonstrated that reciprocity can be broken in photonic devices by combining Kerr-type optical nonlinearity of the signal with spatial asymmetry of the structure~\cite{Fan2012,Yang2019PRL,Yang2020}. Although this approach allows passive asymmetric transmittance for forward and backward propagating light without any external bias, it is prevented from the realization of ideal optical isolation~\cite{Shi2015} because the isolation of noise and the transmission of signal cannot be simultaneously achieved. Alternatively, ideal optical non-reciprocal phase shifting and isolation are possible by directional spatial-temporal modulation of the optical medium~\cite{Asadchy2020,Sounas2017,Yu2009}. Utilizing the nonlinear optical wave-mixing for coherent frequency conversion, a linear non-reciprocal susceptibility of the system with respect to the input signal can be realized by external drive fields. For instance, by either optical~\cite{Shen2016,Zhang2018,Abdelsalam2020,Liang2020}, acoustic~\cite{Tian2021,Sohn2021,Kittlaus2021} or microwave~\cite{Lira2012,Dostart2021} drives, directional conversion between the input signal and idle light can be realized. Since the underlying mechanism could be attributed to the phase-matching condition between traveling waves, this approach not only requires the drive fields and input signal in certain spatial modes but also imposes challenges in separating the signal from the drive field and idle outputs. %when using resonant-enhanced interaction. Thus the performances are limited.

Here, we propose and demonstrate the nonlinear non-reciprocal susceptibility of the optical medium, which allows the ideal optical isolation with neither the requirements of an external bias field nor the phase-matching condition. The time-reversal symmetry of the optical system is broken by the input signal itself, and the non-reciprocity property of the medium is sustained by the transmitting signal, while the counter-propagating light is blocked simultaneously. Remarkably, self-induced isolation with a $63.4\,\mathrm{dB}$ isolation ratio and a large bandwidth has been demonstrated. The self-induced non-reciprocity also brings novel functional optical devices, such as circular polarization purification and cavity-induced isolation through the non-reciprocal leverage effect. Our demonstration unveils a new approach for realizing a practical compact and high-performance isolator by the very simple structure, and such a mechanism is applicable to rare-earth atom-doped optical waveguides for passive and integrated non-reciprocal devices.

\noindent
\begin{figure}
\begin{centering}
\includegraphics[width=1\columnwidth]{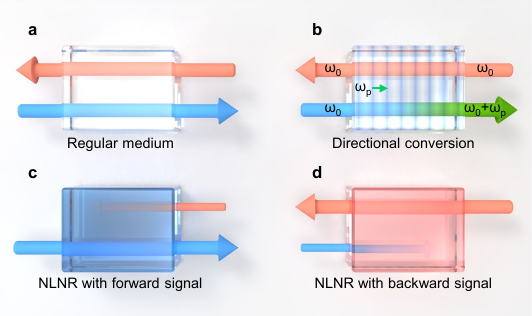}
\par\end{centering}
\caption{\textbf{Schematic diagram of reciprocal and non-reciprocal optical media.} \textbf{a}, The regular medium that is transparent for both forward (blue arrow) and backward (red arrow) propagating light. \textbf{b}, The medium under spatial-temporal modulation due to an external drive ($\omega_{\mathrm{p}}$). The non-reciprocity is induced by the directional coherent conversion ($\omega_{0}\rightarrow \omega_0+\omega_{\mathrm{p}}$) for the forward signal. \textbf{c},\textbf{d}, Nonlinear non-reciprocal (NLNR) medium. The input signal induces non-reciprocal responses of the medium, so the direction of the isolation could be switched when changing the direction of the input signal.}
\label{Fig1}
\end{figure}

%\newpage
\begin{figure*}[t]
\begin{centering}
\includegraphics[width=1\textwidth]{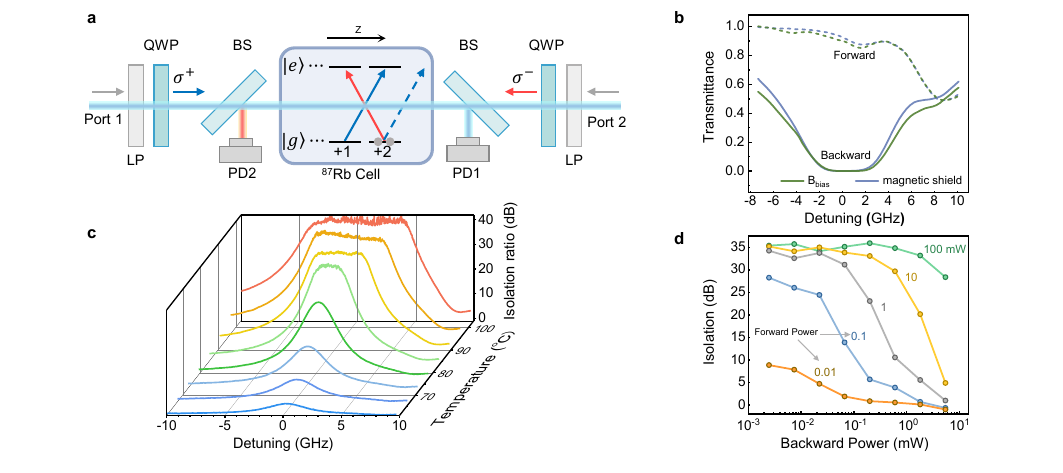}
\par\end{centering}
\caption{\textbf{Experimental setup and characterization of the isolation capability.}
\textbf{a}, Schematic of the experimental apparatus. LP: linear polarizer, QWP: quarter wave plate, BS: beam spillter, PD: photo detector. The kernel device of self-induced non-reciprocity is composed of a $10\,\mathrm{mm}$ Rb vapor cell filled with buffer gas, two LPs and two QWPs. The inset on the vapor cell denotes the energy structure of $^{87}$Rb, with the energy levels $\ket{g}$ and $\ket{e}$ denoted 5$^{2}S_{1/2}$ $F=2$ and 5$^{2}P_{1/2}$ $F=2$, respectively. Blue and red arrows represent the regulated $\sigma^{+}$ and $\sigma^{-}$ polarization of the froward and backward light, respectively. \textbf{b}, Forward $\sigma^{+}$ and backward $\sigma^{-}$ transmission at $81\,^{\circ}\mathrm{C}$ under two circumstances: applying a $5\,\mathrm{Gauss}$ bias magnetic field or using a magnetic shield. \textbf{c}, Isolation spectra under different temperatures. The highest isolation $39\,\mathrm{dB}$ is reached when the temperature is higher than $93\,^{\circ}\mathrm{C}$, and a $12.5\,\mathrm{GHz}$ bandwidth for $20\,\mathrm{dB}$ isolation is realized at $103\,^{\circ}\mathrm{C}$. For the results in both \textbf{b} and \textbf{c}, the forward power is $100\,\mathrm{mW}$, and the backward power is $10\,\mu\mathrm{W}$. \textbf{d}, Maximum isolation ratio under different forward and backward powers at $84\,^{\circ}\mathrm{C}$. Colored numbers beside the lines represent the forward power: 0.01, 0.1, 1, 10, $100\,\mathrm{mW}$.}
\label{Fig2}
\end{figure*}

\begin{figure*}[t]
\begin{centering}
\includegraphics[width=1\textwidth]{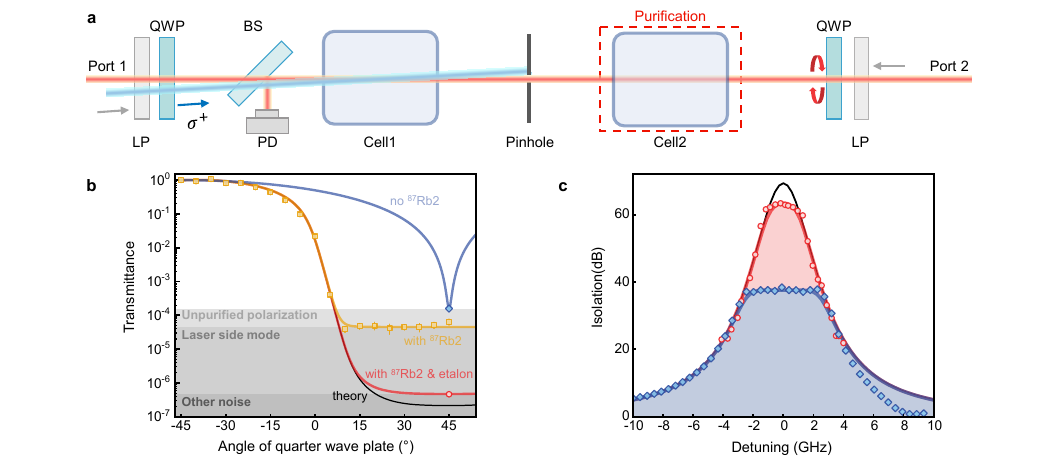}
\par\end{centering}
\caption{\textbf{Ultrahigh isolation via optical circular-polarization purification.} \textbf{a}, Schematic of the improved experimental apparatus with an extra Rb vapor cell. The backward probe purified through Cell2 was used to characterize the isolation ratio in Cell1. \textbf{b}, Transmittance of the backward probe against its polarization, which is controlled by the angle of the QWP (near port 2) with a forward signal power of $150\,\mathrm{mW}$ and a backward probe power of $1\,\mathrm{mW}$. The zero angle corresponds to a linear polarization. The four lines show the corresponding theoretical predictions under different conditions, while the dots are the experimental results. Shaded areas denote noise floors from different causes. \textbf{c}, The improved measurement of the isolation ratio (red circles) by using the NLNR effect for circular-polarization purification and an etalon to eliminate laser background noise, compared to the results without purification (blue diamonds). These two results correspond to $\theta=45^{\circ}$ in \textbf{b}. The highest isolation ratio reaches $63.4\,\mathrm{dB}$ with a $2.1\,\mathrm{GHz}$ bandwidth for $60\,\mathrm{dB}$ isolation. The black line is the theoretical prediction of the ideal isolation ratio, while the red and blue lines are the results considering two different noise floors to fit the experimental data.}
\label{Fig3}
\end{figure*}

\begin{figure*}[t]
\begin{centering}
\includegraphics[width=1\textwidth]{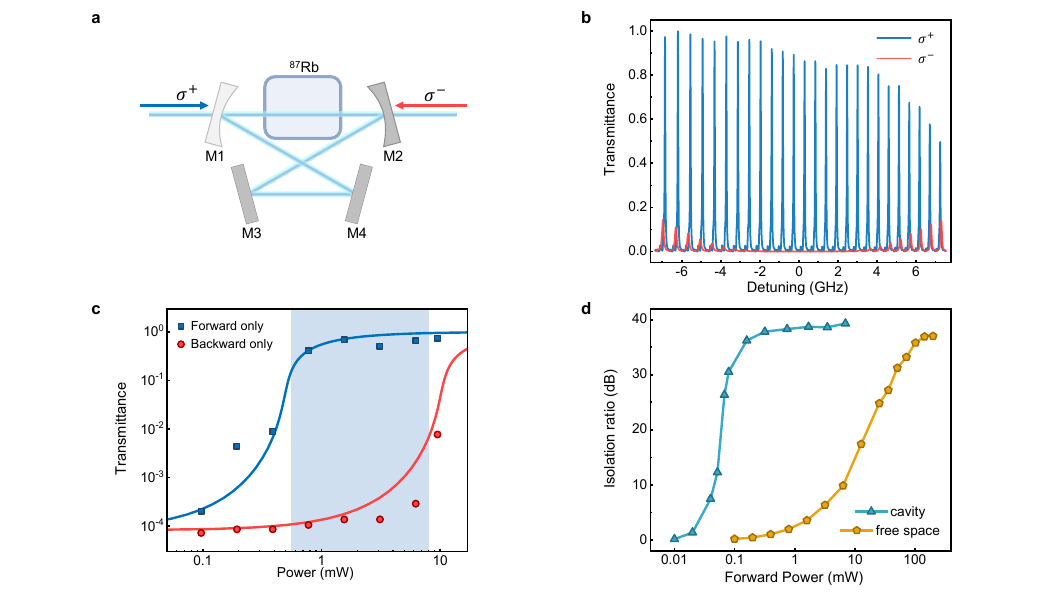}
\par\end{centering}
\caption{\textbf{Cavity-induced non-reciprocal leverage.} \textbf{a}, The experimental apparatus of an asymmetric traveling wave cavity, comprising four mirrors and an Rb vapor cell inside. \textbf{b}, Backward transmission spectra for the $\sigma^{+}$ and $\sigma^{-}$ probes when the cavity is resonantly driven by a forward signal laser. The powers of the forward laser and backward laser are both $7\,\mathrm{mW}$. \textbf{c}, Forward and backward transmission when only one laser is on. Solid lines are the theoretical prediction while dots correspond to the measured results. The blue shadow area indicates the power range where the cavity-induced isolation of the backward probe is effective even if there is no forward signal. \textbf{d}, A contrast of the dependence of isolation on the forward signal power for cavity and free space configurations. The backward laser power is $7\,\mathrm{mW}$ for both lines.}
\label{Fig4}
\end{figure*}

\smallskip{}

\noindent \textbf{Principle}

\noindent Figure$\,$\ref{Fig1} illustrates the reciprocal and non-reciprocal optical media. In Fig.$\,$\ref{Fig1}a, the response of a regular medium to optical light, i.e., the linear susceptibility, is symmetric for forward and backward propagating beams. When utilizing the nonlinear optical susceptibility of the medium, an external directional driving field could induce the spatial-temporal modulation of the refractive index (Fig.$\,$\ref{Fig1}b). Due to the phase-matching condition, the time-reversal symmetry is broken because the forward propagating signal is converted to idle frequency, while the backward propagating signal is free from nonlinear frequency conversion. In practice, the performance of this approach is limited by the imperfect conversion efficiency, and the bandwidth is limited by the strict phase-matching condition.
%requires the strict design of the structure, and consequently the frequency matching and drives, also limited isolation ratio due to imperfection in practical frequency converters. and the corresponding conversion efficiency and bandwidth are limited.

Figures$\,$\ref{Fig1}c and d explain the proposed optical self-induced nonlinear non-reciprocal (NLNR) medium. When there is a forward propagating signal, the components of susceptibility for circular birefringence and circular dichroism effects can be induced by the signal. Such an NLNR medium could be realized by atom ensembles, as the population of the atomic degenerated Zeeman ground sublevels could be polarized by the circularly polarized ($\sigma^{\pm}$) input light~\cite{Hu2021}. Then, the response of such an optical medium to the input field ($\mathbf{E}$) propagating along the $z$-direction ($\mathbf{e}_z$) can be written as~\cite{boyd2020}
\begin{equation}
    \left(\begin{array}{c}
    P_x\\
    P_y\\
    \end{array}\right)=\left(\begin{array}{cc}
    \chi_{xx} & i\chi_{xy}\\
    -i\chi_{xy} & \chi_{yy}\\
    \end{array}\right)\left(\begin{array}{c}
    E_x\\
    E_y\\
\end{array}\right).
\end{equation}
If $\chi_{xy}\neq0$, the medium is non-reciprocal, as their optical susceptibility is anti-symmetric under time reversion, i.e., the refractive index for a $\sigma^{\pm}$-polarized forward propagating light is different from that for a  $\sigma^{\mp}$-polarized backward propagating light (see Supplementary Information for details). Since such a non-reciprocal response is induced by the input, the non-reciprocal susceptibility could be expanded as
\begin{equation}
    \chi_{xy}=\chi_{xy}^{(1)}\mathbf{B}\cdot\mathbf{e}_z+\chi_{xy}^{(3)}(\mathbf{E}\times\mathbf{E}^*)\cdot\mathbf{e}_z+...
\end{equation}
Here, the first term corresponds to the conventional linear non-reciprocal susceptibility for the signal due to the magneto-optics effect, with $\mathbf{B}$ denoting the external bias magnetic field or the effective magnetic field induced by external drives. Other terms denote the nonlinear part of the susceptibility, and here we focus on the second term that corresponds to the third-order NLNR susceptibility, which depends on the local spin property $(\mathbf{E}\times\mathbf{E}^*)$ of the optical field~\cite{Lodahl2017}. Considering the circular dichroism ($\mathrm{Im}(\chi_{xy}$)) of the NLNR material with thickness $L$, the corresponding power transmittance of the counter-propagating noise compared to that of the input is $e^{-4k\mathrm{Im}(\chi_{xy}) L}$ with the signal wave vector $k$, which leads to isolation. Comparing Fig.$\,$\ref{Fig1}c and d, the isolation by NLNR is reconfigurable when changing the direction of the input. Similarly, the signal-induced circular birefringence $\mathrm{Re}(\chi_{xy})$ would lead to a non-reciprocal phase $\phi=2k\mathrm{Re}(\chi_{xy}) L$, which enables the optical gyrator and circulator. Note that the non-reciprocal susceptibility is the intrinsic response of each part of the medium, in contrast to the interaction between certain spatial modes for the modulation approach (Fig.$\,$\ref{Fig1}b).

\smallskip{}

\noindent \textbf{Self-induced isolation}

\noindent The NLNR is experimentally demonstrated with a $^{87}$Rb atom ensemble in a Rb vapor cell filled with nitrogen buffer gas ($200\,$Torr). As schematically shown by the experimental apparatus in Fig.$\,$\ref{Fig2}a, the two ports (1 and 2) are regulated by linear polarizers and quarter-wave plates (QWPs). The forward input laser from Port 1 is $\sigma^{+}$-polarized and reconfigures the NLNR medium by coupling to the D1 transitions of the $^{87}$Rb atoms ($795\,$nm). Therefore, the backward input at Port 2 could be converted to $\sigma^{-}$ and be absorbed by the atoms. It's worth noticing that a backward propagating light with any polarization cannot pass through the isolator because it will either be absorbed by the atom or blocked by the polarizer, which is similar to the traditional Faraday isolator. As shown by the energy diagram, at the microscopic level of each atom, the population of ground sublevels is polarized by the circular polarization input and thus leads to polarization-dependent interactions with photons.

Figure$\,$\ref{Fig2}b demonstrates the non-reciprocal circular dichroism response of the medium by sending the signal transmitting forward through the system and simultaneously measuring the transmission of a backward probe from another laser with a similar frequency. The distinct contrast of the spectra for the forward and backward directions shows the excellent isolation of the backward probe over a broad frequency range, with the insertion loss of the forward input being only $0.5\,\mathrm{dB}$. Here, to avoid the environment magnetic field-induced depolarization of the atom ground state populations, the cell is placed in a magnetic shield. Therefore, a completely passive NLNR is demonstrated without any external bias field or drive lasers.

For a concise investigation of the NLNR under different experimental conditions, we apply a very weak magnetic field $B_{\mathrm{bias}}\approx5\,\mathrm{Gauss}$ along the $z$-direction instead of the shield to simply the setup (Fig.$\,$\ref{Fig2}a). Noting that weak $B_{\mathrm{bias}}$ is too weak to directly induce any observable non-reciprocity in our system, it could effectively protect the NLNR medium from stray magnetic fields, as shown by the spectra in Fig.$\,$\ref{Fig2}b (see Supplementary Information for more details). The performance of the NLNR is characterized by $\mathcal{I}$, which is defined as the ratio of the backward transmission of $\sigma^{+}$ and $\sigma^{-}$. $\mathcal{I}$ is equivalent to the isolation ratio between forward and backward probes, and it holds the advantages in practice that the influence of the differences in optical insertion losses between forward and backward paths is mitigated.
%This approximation is feasible because when the weak backward probe is too weak to change the population of the atoms, the absorption loss for backward $\sigma^{-}$ laser due to the atoms is exact the same as the input signal.
%\Zcb{Comparing with conventional non-reciprocal devices, the NLNR medium is transparent to the input signal no matter which direction it comes in, but rejects the weak backward propagating stray light.}

Typical isolation ratios as a function of the probe laser frequency are shown in Fig.$\,$\ref{Fig2}c. The density of the atoms is increased by approximately 2 orders of magnitude by varying the temperature of the cell from $61\,^{\circ}\mathrm{C}$ to $103\,^{\circ}\mathrm{C}$, and thus, the optical depth of the atomic ensemble is boosted. Both the isolation bandwidth for $\mathcal{I}>20\,\mathrm{dB}$ and the maximum isolation ratio increase as the temperature increases. Thanks to the transition broadening due to the buffer gas in the vapor cell, an isolation bandwidth exceeding $12.5\,\mathrm{GHz}$ at $103\,^{\circ}\mathrm{C}$, which is orders of magnitude higher than the previous demonstration in atoms~\cite{Zhang2018,Liang2020,Hu2021}. Another figure-of-merit for the isolation is the insertion loss, which is typically smaller than $~0.5\,\mathrm{dB}$ below $90\,^{\circ}\mathrm{C}$ and increases to $~1.9\,\mathrm{dB}$ under $103\,^{\circ}\mathrm{C}$ (see Supplementary Information for more details).
%For the self-induced non-reciprocity, the working power dynamic range is important for its practical applications.
In Fig.$\,$\ref{Fig2}d, the dependence of $\mathcal{I}$ on the backward probe laser power $P_{b}$ under different forward signal powers $P_{\mathrm{f}}$ is investigated. When the backward probe is strong enough, the NLNR effect due to the probe cannot be ignored, as the probe could cancel or overcome the non-reciprocal susceptibility induced by the signal. Therefore, a general trend of the degradation of $\mathcal{I}$ is observed for larger $P_{b}/P_{f}$.

\smallskip{}
\noindent \textbf{Non-reciprocal purification and leverage}

\noindent The flat-top spectra for the dense NLNR medium shown in Fig.$\,$\ref{Fig2}c indicate a much higher $\mathcal{I}$ around zero detuning hindered by noise. After a thorough inspection of the system, we found that the measured $\mathcal{I}$ is essentially limited by the imperfect polarizer and waveplate (Fig.$\,$\ref{Fig2}a), which generates an impure circular polarization, i.e., the backward probe contains the $\sigma^{+}$ component of a portion around $~10^{-4}$ ($100\,\mathrm{ppm}$). The detected flat-top spectra imply that our NLNR provides an ultrasensitive circular-polarization analyzer and suggests a potential application of the NLNR device for optical circular-polarization purification. As shown by the setup in Fig.$\,$\ref{Fig3}a, we use an additional NLNR medium to purify the polarization of the probe. Specifically, a strong enough ($1\,\mathrm{mW}$) backward $\sigma^{-}$ probe, with a small portion of $\sigma^{+}$ component in it, will polarize the medium in Cell2 to only absorb $\sigma^{+}$ light, and thus purifies itself. Figure$\,$\ref{Fig3}b presents the transmittance of the probe for different input polarizations controlled by the angle of the QWP. Treating the self-induced isolation (left half of the setup) as a black box that only $\sigma^{+}$-polarized light can pass through, we can expect a transmittance of $cos^2(\theta+\frac{\pi}{4})$ without Cell2. However, the NLNR effect in Cell2 gives a significantly different response, as the measured transmittance of the laser is $0.02$ when $\theta=0$, in contrast to $0.5$ from $cos^2(\frac{\pi}{4})$. The observed purification is in excellent agreement with theoretical fitting (see Supplementary Information) and reveals other imperfections due to laser background noise. By significantly improving the circular polarization purity and filtering the background noises, we achieved an ultrahigh isolation ratio of $63.4\,\mathrm{dB}$, with a $60\,\mathrm{dB}$ isolation bandwidth as large as $2.1\,\mathrm{GHz}$. The experimental spectra are consistent with our theoretical model, which predicts a purity of the circular polarization  better than $0.5\,\mathrm{ppm}$, and the achievable isolation is $70\,\mathrm{dB}$.
%,  As a result of the,  \Zcr{Noting that for a better signal-to-noise ratio, the backward light is misaligned with respect to the forward signal, and the results manifests the inherent medium non-reciprocal susceptibility that do not require the phase-matching condition and is robust against the alignment.}

It is anticipated that the direct application of the NLNR might be limited to scenarios where the input signal power is strong enough and the backward light is much weaker. Harnessing the inherent nonlinearity of the effect, this drawback could be mitigated by non-reciprocal leverage. As shown in Fig.$\,$\ref{Fig4}a, by placing the NLNR medium into an asymmetric cavity with mirrors of different reflectivities, the optical isolation could be leveraged from two aspects. First, the intracavity NLNR could be enhanced by the resonance, thus the forward signal power to activate the medium to be transparent is lowered. Second, the asymmetric cavity containing the Kerr-like $\chi^{(3)}_{xy}$ nonlinear medium allows dynamic non-reciprocity~\cite{Fan2012,Yang2019PRL,Yang2020}.
%The isolation effect would be switched only when the intracavity backward propagating light power is orders higher than the forward power. Thirdly, by employing a mirror with higher reflectivity, the asymmetric would further enhance the leverage effect.
Figure$\,$\ref{Fig4}b presents the transmission spectra of the non-reciprocal leverage for individual forward and backward input. The mirror transmittances are 0.082 (M1) and 0.004 (M2), respectively, and the laser power is $7\,\mathrm{mW}$ for both cases. The result shows that the broken spatial symmetry induced the blockade of the backward light even though there is no forward input, in contrast to the free-space case. The shifts between the forward and backward resonances at large detunings are due to the circular birefringence.

The power-dependence relation of the resonance transmittance around zero detuning for both directions is tested in Fig.$\,$\ref{Fig4}c. It is found that leverage could be activated when $P_{\mathrm{F}}>400\,\mathrm{\mu W}$, but the backward light is blocked as long as its power $P_{\mathrm{B}}<8\,\mathrm{mW}$. The non-reciprocal leverage effect is manifested by testing the isolation $\mathcal{I}$ when the non-reciprocal leverage is sustained by an input $P_{\mathrm{F}}$, with a fixed backword $P_{\mathrm{B}}=7\,\mathrm{mW}$. We found that $\mathcal{I}$ could be as large as $30\,\mathrm{dB}$ even when $P_{\mathrm{F}}=0.01P_\mathrm{B}$, as shown in Fig.$\,$\ref{Fig4}d.
%When applying a continuous signal to the leverage with various $P_{\mathrm{F}}$ and testing the isolation of backward light with $P_{\mathrm{B}}=7\,\mathrm{mW}$, We found that the  isolation ratio could be as larger as $30\,\mathrm{dB}$ even when $P_{\mathrm{F}}=0.01P_\mathrm{B}$, as shown in Fig.~\ref{Fig4}D.
Compared with the NLNR medium in free space, the leverage shows a 3 orders of magnitude improvement on the requirement of $P_{\mathrm{F}}$ for $\mathcal{I}=30\,\mathrm{dB}$.

\smallskip{}

\noindent \textbf{\large{}Discussion}{\large\par}

\noindent Our experiments accomplish a new route for realizing functional and high-performance non-reciprocal photonic devices. By exploiting the nonlinear non-reciprocal susceptibility of the optical medium, the non-reciprocity could be realized by the input signal itself and leveraged by spatial asymmetry, without requirements for any external driving field or a strict phase-matching condition for the input. Benefiting from the combined nonlinearity and non-reciprocity, novel photonic devices, such as circular polarization filters, are enabled. Notably, the principle of NLNR validated in rubidium gas can be easily extended to other atoms and molecules for non-reciprocity at UV, mid-infrared or THz frequencies and thus could also be implemented by integrated photonic structures doped with rare-earth atoms. In the future, our setup could be minimized to realize compact, passive, and high-performance devices, which could replace commercial products when using semiconductor lasers in the studies of atomic and molecular physics. Furthermore, the inherent non-reciprocal nonlinear susceptibility could be generalized to other nonlinear effects, such as the cross-Kerr and coherent frequency mixing processes. Therefore, this work presents a significant conceptual advance in optics and could stimulate further exploration of physics with nonlinear dynamics and non-reciprocity by considering the interplay between the optical fields, microwave fields, and configuration (internal state population) of the medium.

\smallskip{}

\noindent \textbf{\large{}Acknowledgment}{\large\par}

\noindent We thank Dong Sheng for fabricating the vapor cells and helpful discussions. This work was supported by the National Key Research and Development Program of China (Grant No.~2021YFA1402004) and the National Natural Science Foundation of China (Grant No.~U21A20433, 11874342, 11934012 and 11922411). This work is also supported by the Natural Science Foundation of Anhui Province (Grant No. 2008085QA34 and 2108085MA22) and the Fundamental Research Funds for the Central Universities. This work was partially carried out at the USTC Center for Micro and Nanoscale Research and Fabrication.

\noindent \textbf{\large{}Methods}{\large\par}

\noindent \textbf{Experimental setup}

\noindent The detailed experimental setup for the self-induced optical non-reciprocity for the cavity-free scenario is illustrated in Extended Data Fig.~\ref{FigEx1}. For the study of cavity-induced leverage, a bow-tie cavity is added to the system by inserting four mirrors around the Rubidium ($^{87}$Rb) vapor cell, with the bow-tie cavity sketched in Fig.~4a in the main text. In the setup, there are two lasers: Laser1 (Toptica DLpro $795\,\mathrm{nm}$) provides the backward probe, which passes through the vapor cell with a beam waist of $600\,\mathrm{\mu m}$, and Laser2 (Wavicle ECDL $795\,\mathrm{nm}$) provides the forward signal with a beam waist of $750\,\mathrm{\mu m}$. Two quarter wave plates (QWP) have two mutually perpendicular fast axes, thus the forward and backward light have orthogonal circular polarization ($\sigma^{\pm}$) when interacting with the atoms in the vapor cell. Meanwhile, the forward (backward) horizontally polarized light can travel from port1(2) to port2(1) without reflection on any polarization beam splitter. Therefore, the setup allows the measurement of the non-reciprocity properties of our devices. More details about the characterization of the non-reciprocity are provided in the following method and Extended Data Fig.~\ref{FigEx2}.

In our experiments, the frequencies of the two lasers are  tuned independently and are near resonance with the D1 transitions of the rubidium atoms (detuning less than $10\,\mathrm{GHz}$). For the measurement of the forward and backward transmittance of light, there is potential cross-talk between the lights that induces difficulties in measuring ultrahigh isolation ratios due to the backgrounds. Therefore, the lock-in amplifier technique is employed in these measurements. We use a chopper (Thorlabs MC2000) and a lock-in instrument (Zurich MFLI 500~kHz) to suppress the background noise and electrical noise on the optical detectors (Thorlabs PDA36A2 \& APD410A).

The atomic vapor cell we used is a $10\,\mathrm{mm}$ cube filled with $^{87}$Rb atoms and $0.23\,\mathrm{amg}$ ($200\,\mathrm{Torr}$ at room temperature) N$_{2}$ as the buffer gas. The buffer gas is beneficial for our device from two aspects. First, it changes the motion of $^{87}$Rb atoms from linear motion to Brownian motion, thus the depolarization of the $^{87}$Rb atoms is suppressed. Additionally, the transient effect is suppressed, and the atoms can be polarized more efficiently. Second, the collision between excited $^{87}$Rb atoms and N$_{2}$ molecules will greatly increase the absorption linewidth, which will improve the isolation bandwidth ($\sim 10\,\mathrm{GHz}$) when compared to the case without buffer gas ($\sim 0.7\,\mathrm{GHz}$ due to Doppler broadening).

Since an external magnetic field would induce the depolarization of the ground states of the atoms, the performance of the self-induced non-reciprocity degrades by the background magnetic fields. Therefore, two different approaches are applied in our experiments to mitigate the influence of the background magnetic field: (1) place the atomic vapor cell inside a magnetic shield, which is a Permalloy cylinder (thickness of $0.5\,\mathrm{mm}$), or (2) apply a very weak bias field along the propagation direction of the signal. By the first approach, the magnetic-free nature of the self-induced non-reciprocity mechanism is experimentally demonstrated, as shown by Fig.~2b in the main text. We found a slight difference between the transmission under the magnetic shield and bias field through Fig.~2b. The transmission rates of forward $\sigma^{+}$ signals are almost the same under different temperatures, which provides additional evidence that the only effect of $B_{\mathrm{bias}}=5\,\mathrm{G}$ is to eliminate stray magnetic fields in the environment, not to induce non-reciprocity. Since the magnetic shield brings difficulties when changing the working temperature of the vapor cell and when working at different optical configurations, we carry out further systematic experimental characterizations of the device under different conditions with a bias magnetic field for convenience.

\vbox{}

\noindent \textbf{Characterization of the isolation ratio}

\noindent The isolation ratio is an important quantity for practical applications of non-reciprocal devices. Usually, the isolation ratio ($\mathcal{I}$) is defined as the ratio between the forward transmittance and the backward transmittance of light. The detailed explanation of the $\mathcal{I}$ is illustrated in Extended Data Fig.~\ref{FigEx2}. Extended Data figure~\ref{FigEx2}a explains the principle of our device when it is applied in practical optical isolation. The device consists of two PBSs, two QWPs ($+45^{\circ}$ and $-45^{\circ}$), and a vapor cell. Note that the device works for H-polarized inputs, and V-polarized light is rejected by the PBS on both ports. When H-polarized light is input from the left, its polarization is converted to $\sigma^{+}$ by the left QWP, and the transmitted light is converted to H-polarization again by the QWP at the output port. When H-polarized light is input from the right, it is converted to $\sigma^{-}$-polarization to couple with the atoms and then output from the left port by converting the light back to H-polarization. Comparing the evolution of forward and backward light propagation, we found that the time-reversal evolution is broken if the atoms show circular dichroism or birefringence.

When measuring the transmittance of light, we could place BS into the optical paths to measure the light intensity. The BSs are placed just close to the cell because the atomic medium provides the essential ingredients for realizing non-reciprocity and because the polarization of forward and backward light are orthogonal and the possible reflections of input signal and noises introduced by other optical components could be reduced.(as shown in
Extended Data Fig.~2a) Therefore, close to the vapor cell, the transmittance of forward $\sigma^{+}$-polarized light ($T_{\mathrm{f,+}}$) and the transmittance of backward $\sigma^{-}$-polarized light ($T_{\mathrm{b,-}}$) could be measured separately by two PDs, and the system isolation ratio could be derived as
\begin{equation}
\mathcal{I}_{\mathrm{sys}}=10\times\mathrm{Log}_{10}\left(\frac{T_{\mathrm{f},+}}{T_{\mathrm{b},-}}\right).
\end{equation}

However, when characterizing the transmittance of light in real experiments, the optical paths and the optical components (including the BSs and detectors) are different for forward and backward light. Therefore, the different optical paths and components could introduce different losses to the transmittance, which is difficult for calibrations of transmittance and eventually causes errors in the estimated $\mathcal{I}$.

For a more precise characterization of the $\mathcal{I}$ and to avoid the difficulties in calibrations, we characterize the $\mathcal{I}$ by measuring the ratio between the transmittance for backward propagating probe light with different polarizations ($\sigma^{+}$- or $\sigma^{-}$-polarized light when passing through the cell), as shown in Extended Data Figs.~\ref{FigEx2}b and c. Here, we adapt the property of the atomic medium that the circular dichroism or birefringence property of the vapor cell is the same for both forward and backward light, i.e., the transmittance of the $\sigma^{+}$-polarized light through the cell should be exactly the same for both the forward and backward directions. Therefore, $T_{\mathrm{f,+}}$ can be obtained by equivalently measuring $T_{\mathrm{b,+}}$ in Extended Data Fig.~\ref{FigEx2}c. The optical configuration for measuring $T_{\mathrm{b,-}}$ (Extended Data Fig.~\ref{FigEx2}b) and $T_{\mathrm{b,+}}$ are exactly the same except that the angle of the QWP is rotated by $90^{\circ}$. Therefore, the potential calibration errors are significantly suppressed when measuring transmittance light by sharing the same optical path and photodetector, and the corresponding

\smallskip{}

\setcounter{figure}{0}

\renewcommand{\figurename}{\textbf{Extended Data Fig.}}
%\captionsetup[figure]{labelfont={bf},labelformat={default},labelsep=period,name={Extended Data Fig.},singlelinecheck=false}

\begin{figure*}[t]
%\begin{centering}
\includegraphics[width=0.8\textwidth]{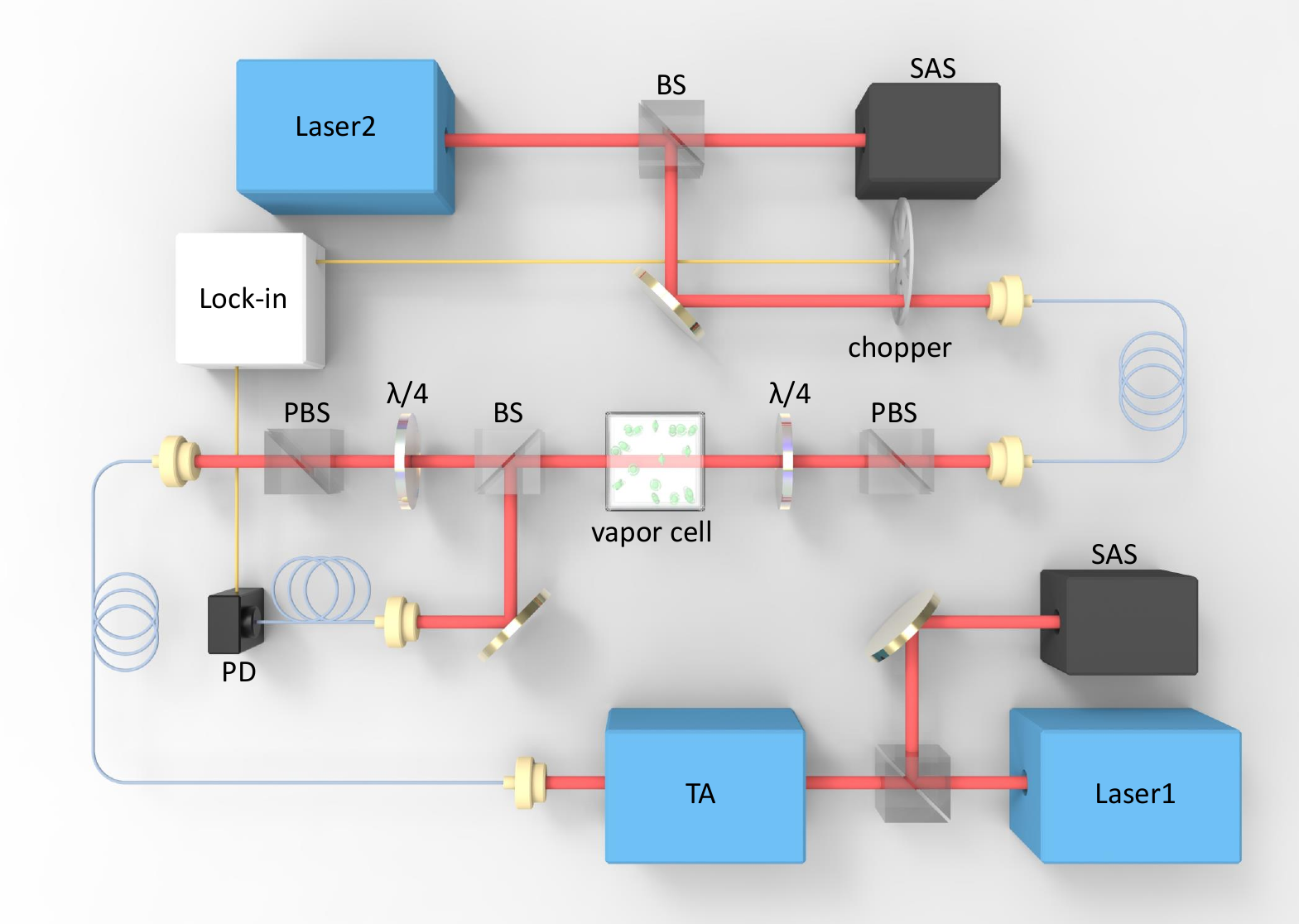}
%\par\end{centering}
\caption{\textbf{Schematic of the experimental setup for the isolation measurement without a cavity.} The red beams denote the free-space optical paths of both the forward signal and backward probe, the gray lines represent the optical fibers, and the yellow lines are electric cables. SAS: saturated absorption spectrum. TA: tapered amplifier. PBS: polarization beam splitter. BS: beam splitter. QWP: quarter wave plate.}
\label{FigEx1}
\end{figure*}

\clearpage{}

\begin{figure*}[t]
%\begin{centering}
\includegraphics[width=0.8\textwidth]{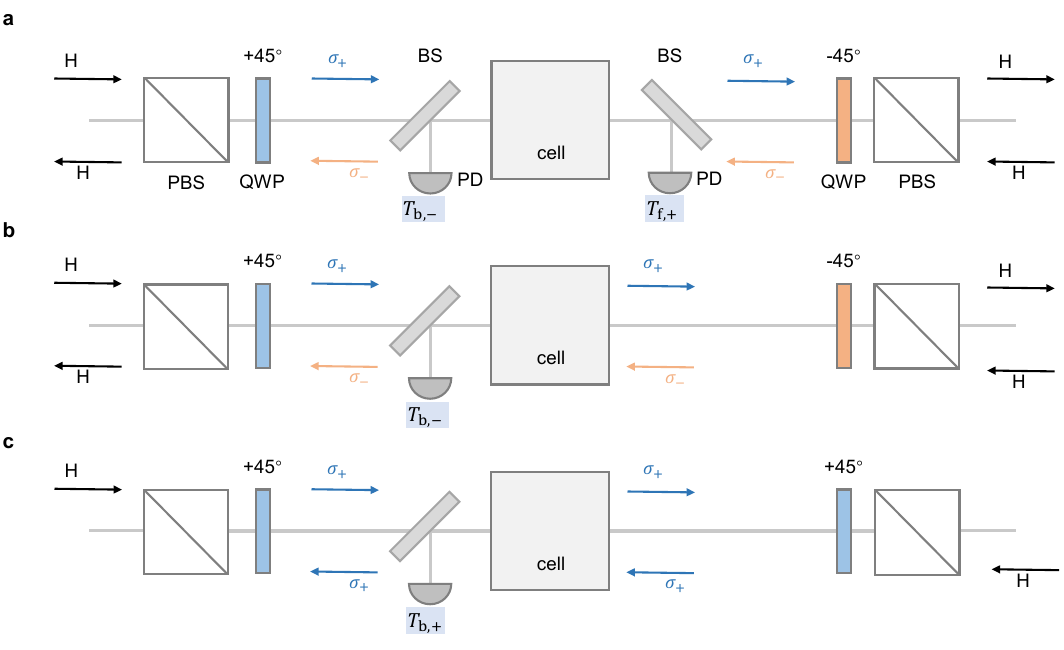}
%\par\end{centering}
\caption{\textbf{Definition of the isolation ratio.} \textbf{a}, The system isolation ratio is defined as the ratio between the forward transmittance and the backward transmittance. $T_{\mathrm{f(b)},\pm}$ is the transmission of the forward (backward) laser with an H-polarization from the input and output ports. The polarization of the light when coupled with the atoms depends on the incident direction due to the quarter wave plates. \textbf{b},\textbf{c}, The experimentally measured isolation ratio is defined as the transmittance ratio between the backward and forward probes that are $\sigma^{+}$-polarized and $\sigma^{-}$-polarized when coupled with the atoms.}
\label{FigEx2}
\end{figure*}

\end{document}